\documentclass[letterpaper]{article}
\usepackage[preprint]{aaai2027}
\usepackage[hyphens]{url}
\usepackage{graphicx}
\usepackage{natbib}
\usepackage{booktabs}
\usepackage{multirow}
\usepackage{amsmath}
\usepackage{amssymb}
\usepackage{placeins}
\usepackage{algorithm}
\usepackage{algpseudocode}
\title{Requirement--Evidence Alignment
for Compositional E-Commerce Queries}
\author{
    Weihao Shen\textsuperscript{\rm 1},
    Wei Chen\textsuperscript{\rm 1},
    Fuwei Zhang\textsuperscript{\rm 1},
    Meng Yuan\textsuperscript{\rm 1},
    Yuqin Lan\textsuperscript{\rm 1},\\
    Guojun Liu\textsuperscript{\rm 2},
    Qingsong Hua\textsuperscript{\rm 2},
    Wei Lin\textsuperscript{\rm 2},
    Fuzhen Zhuang\textsuperscript{\rm 1}\corresponding
}
\affiliations{
    \textsuperscript{\rm 1}Institute of Artificial Intelligence, Beihang University, Beijing, China\\
    \textsuperscript{\rm 2}Meituan, Beijing, China
}

\begin{document}

\maketitle

\begin{abstract}
Compositional e-commerce queries express multiple requirements that must hold jointly, yet existing rerankers collapse these constraints into aggregate relevance and often promote topical near misses over feasible products. In this paper, we introduce \textbf{REAlign}, a novel requirement–evidence-aligned reranking framework that explicitly connects typed query requirements with visible evidence. REAlign distinguishes satisfied, violated, and unsupported conditions, constructs requirement-targeted contrasts that expose failure modes, and optimizes duplicate-free partial rankings through Requirement-Aware Group-Relative Policy Optimization. Its list utility preserves relevance while incorporating requirement satisfaction, evidence support, material violations, and output validity. Experiments on two fixed-pool e-commerce benchmarks show consistent improvements over strong supervised and policy-optimization baselines under matched training budgets, with fewer violations among top-ranked candidates and larger gains at shallow ranks. Controlled ablations confirm the complementary value of requirement modeling, evidence grounding, and decomposed optimization.
\end{abstract}
\begin{links}
\link{Code}{https://github.com/Nevaeh7/REAlign}
\end{links}
\section{Introduction}
\label{sec:introduction}

E-commerce search systems commonly adopt a retrieval-and-reranking
pipeline. Given a user query and a candidate pool produced by an
upstream retriever, a reranker reorders the candidates so that the
products most useful to the shopper appear at the top~\citep{puthenputhussery2025walmart,sheng2025progressive,chen2026walmartcrossencoder}. Existing
rerankers typically formulate this task in terms of query--product
relevance and are trained using pairwise relevance
signals. This formulation is effective when a query primarily expresses
a topical intent, such as a product category or brand.

\begin{figure}[!t]
\centering
\includegraphics[width=\columnwidth]{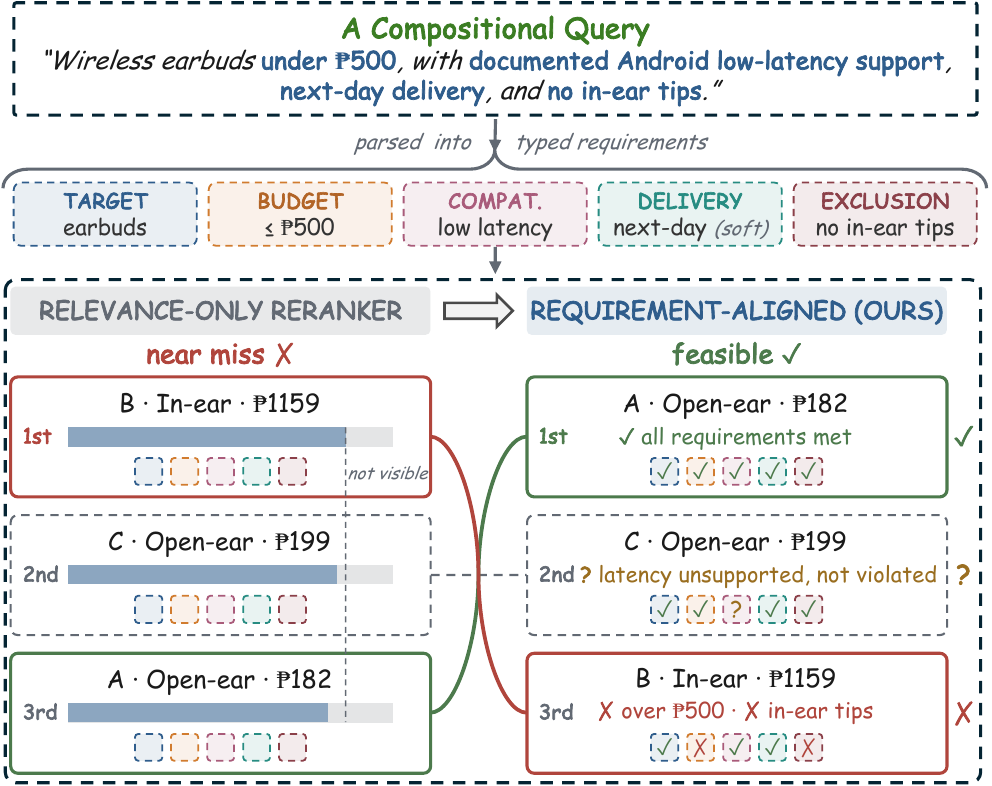}
\caption{
The relevance--requirement gap.
Relevance scores are nearly tied (left; illustrative), so a
relevance-only reranker places near miss B first despite violations.
Requirement--evidence-aligned reranking (right) promotes feasible A
and distinguishes \emph{unsupported} (C) from \emph{violated} (B).
}
\label{fig:intro}
\end{figure}

In practice, many shopping queries are compositional. In addition to
specifying a target product, users may jointly express requirements
concerning product attributes, budgets, exclusions, compatibility, and
usage scenarios~\citep{wang2026shoppingbench,krishnasamy2026constraint,zhang2025excluir}. A useful result must therefore be
not only topically relevant but also compatible with the requested
conditions. Figure~\ref{fig:intro} illustrates this distinction. A user
searches for wireless earbuds under 500 pesos, with documented
low-latency support for Android, next-day delivery, and no in-ear tips.
All candidates are topically relevant and receive nearly tied relevance
scores. Nevertheless, a conventional reranker places an over-budget
candidate with the wrong form factor first, while ranking the only
feasible option last. For compositional queries, topical relevance is
therefore necessary but insufficient.

This failure stems partly from the supervision used to train most
rerankers. Aggregate relevance labels indicate whether a candidate
should rank highly, but do not identify which query requirement the
candidate satisfies, violates, or lacks evidence to support. When
candidates belong to the same category and share similar titles,
descriptions, and marketing language, topical similarity becomes the
easiest signal for a model to exploit. The reranker may consequently
learn a topical shortcut: it rewards broad semantic compatibility
without determining whether the candidate is actually usable under the
shopper's requirements.
We refer to this mismatch between aggregate relevance supervision and
requirement-level decision making as the
\textbf{relevance--requirement gap}. A particularly challenging
consequence is the prevalence of near misses: semantically
plausible candidates that fail one or more material requirements~\citep{elsen2025nevir,alambo2026intent}, such as
candidate B in Figure~\ref{fig:intro}. Because such candidates remain
highly relevant at the topical level, conventional relevance
supervision provides only a weak signal for distinguishing them from
feasible products.

Existing approaches address individual components of this problem.
Multi-aspect representations and query decomposition expose
fine-grained shopping intent
\cite{kong2022multiaspect,luo2023implicit,sun2024mural},
while structured catalog representations organize product attributes
and relations
\cite{zhu2026attributegraph}.
Hard-negative mining identifies semantically challenging candidates
\cite{xiong2021ance,qu2021rocketqa,zhou2022simans},
but typically does not specify the particular requirement responsible
for a candidate's failure. Listwise language-model rerankers compare
multiple candidates jointly
\cite{sun2023rankgpt,reddy2024first,liu2026reasonrank,ren2025selfcalibrated},
and recent reinforcement-learning methods directly optimize ranking
behavior
\cite{zhang2025rearank,li2026prorank,long2026grouprank,wu2026rrpo}.
These components, however, remain largely disconnected. Fine-grained
query requirements are not systematically aligned with visible
candidate evidence, difficult negatives are not organized by
requirement-specific failure roles, and list-level objectives do not
explicitly distinguish feasible candidates from near misses.

Addressing this gap requires more than decomposing a long query into a
collection of attributes. A reranker must connect each requested
condition to candidate-side evidence and determine whether that
condition is satisfied, violated, or unsupported, as illustrated by
candidate C in Figure~\ref{fig:intro}. It must further identify
why a topically plausible candidate should be demoted and make
requirement-level differences consequential to the final ranking
objective. In particular, candidates that fail different requirements
should not be treated as interchangeable hard negatives.

To address this gap, we propose \textbf{REAlign}, which aligns typed query requirements
with visible candidate evidence and constructs requirement-specific contrasts
among similar candidates.
Given a fixed candidate pool, REAlign optimizes duplicate-free partial
top-$K$ rankings with Requirement-Aware Group-Relative Policy
Optimization (R-GRPO). Its list-level objective preserves source
relevance while rewarding supported requirement satisfaction and
penalizing material violations and invalid outputs. Unlike hard constraint
filtering, REAlign treats missing evidence as unresolved and distinguishes
hard requirements from soft preferences, allowing trade-offs under incomplete
catalog information.

We evaluate REAlign on two constructed fixed-pool e-commerce reranking
benchmarks, \textsc{Shop-Need} and \textsc{KS-Need}. Under matched
training budgets, REAlign consistently outperforms relevance-oriented
GRPO and standard policy-optimization baselines, with particularly
strong gains at shallow ranks and on more compositional queries. It
also reduces top-ranked constraint violations without sacrificing
evidence support. Controlled ablations further confirm the
complementary contributions of requirement modeling, evidence
grounding, and requirement-aware optimization.

Our main contributions are summarized as follows:
\begin{itemize}
    \item
    We identify the relevance--requirement gap in compositional
    e-commerce reranking: aggregate relevance supervision cannot
    reliably separate feasible products from topically relevant
    near misses.

    \item
    We introduce REAlign, a new supervision and optimization framework
    that connects typed query requirements to visible candidate
    evidence, constructs requirement-specific contrasts, and optimizes
    duplicate-free partial rankings with R-GRPO.

    \item
    We provide evidence on two fixed-pool benchmarks that
    requirement--evidence alignment improves ranking effectiveness,
    reduces violations, and becomes increasingly valuable
    as query complexity grows.
\end{itemize}
\section{Related Work}

\paragraph{Compositional intent in e-commerce search.}
Sparse and neural rankers learn topical or behavioral relevance from text and
interaction signals
\citep{robertson2009probabilistic,nguyen2020robust}.
Multi-Aspect Dense Retrieval, MURAL, and implicit query parsing represent
shopping intent at finer granularity
\citep{kong2022multiaspect,sun2024mural,luo2023implicit,sheng2025progressive}. Decomposition hints
and explicit intent modeling further improve query--item comparison
\citep{alambo2026intent,luo2025shoppingalignment}. Recent work tracks fulfillment
intent and persistent constraints in product search
\citep{xu2025fulfillment,krishnasamy2026constraint}. REAlign addresses the
downstream supervision interface: typed requirements are tied to visible
evidence, requirement-specific failures, and an ordered subset over a
fixed candidate pool.

\paragraph{Catalog structure and evidence.}
Attributes, entities, and structured relations are established resources for
retrieval. Centrality-aware ranking exploits catalog relations
\citep{choudhary2022graph}, and LLM-guided attribute graphs structure
unstructured catalog content for search \citep{zhu2026attributegraph}. REAlign
therefore makes no standalone novelty claim for attributes or graphs. It treats
catalog fields as evidence during construction: internal evidence cards
retain sources, while the compact policy receives visible fields
and aggregate alignment features.

\paragraph{Hard negatives and requirement contrasts.}
Dense retrieval often mines close negatives through approximate-neighbor,
denoising, or ambiguity-aware strategies
\citep{xiong2021ance,qu2021rocketqa,zhou2022simans}. Similarity-based hardness,
however, does not necessarily identify why a candidate should lose. REAlign
constructs a targeted near miss by changing or selecting against a designated
requirement such as category, budget, or visible evidence support. We call this a
requirement contrast: a candidate-level data operation, distinct from
counterfactual learning-to-rank from biased logs \citep{joachims2017unbiased}.

\paragraph{LLM ranking with reinforcement learning.}
Large language models have been adapted to ranking through pointwise,
pairwise, listwise, and permutation-generation paradigms
\citep{qin2023pairwise,chao2024alro,chen2025tourrank,ji2025reasontorank}.
Recent studies directly optimize the generated ordering with reinforcement
learning.
REARANK learns from ranking-oriented feedback; ProRank and GroupRank combine
group-relative policy optimization with list-level rewards
\citep{zhang2025rearank,li2026prorank,long2026grouprank}. RRPO introduces
rank-level credit assignment, while Rank-GRPO and F-GRPO develop further
group-relative formulations for recommendation and ranking
\citep{wu2026rrpo,zhu2026rankgrpo,surana2026fgrpo}.
Related studies stabilize and accelerate group-based post-training
\citep{zhang2025gvpo}, and reinforcement learning now
reaches production e-commerce relevance systems \citep{yang2026taosr}.

Despite these advances, existing approaches primarily improve the ranking
protocol or the policy-optimization procedure, while the semantics of the
optimized reward are typically specified through aggregate relevance or
generic list-quality signals.
REAlign closes this gap by making typed requirements, visible evidence, and
requirement-specific failure modes first-class ranking signals.
\section{Methodology}
\label{sec:method}

\begin{figure*}[!t]
\centering
\includegraphics[width=\textwidth]{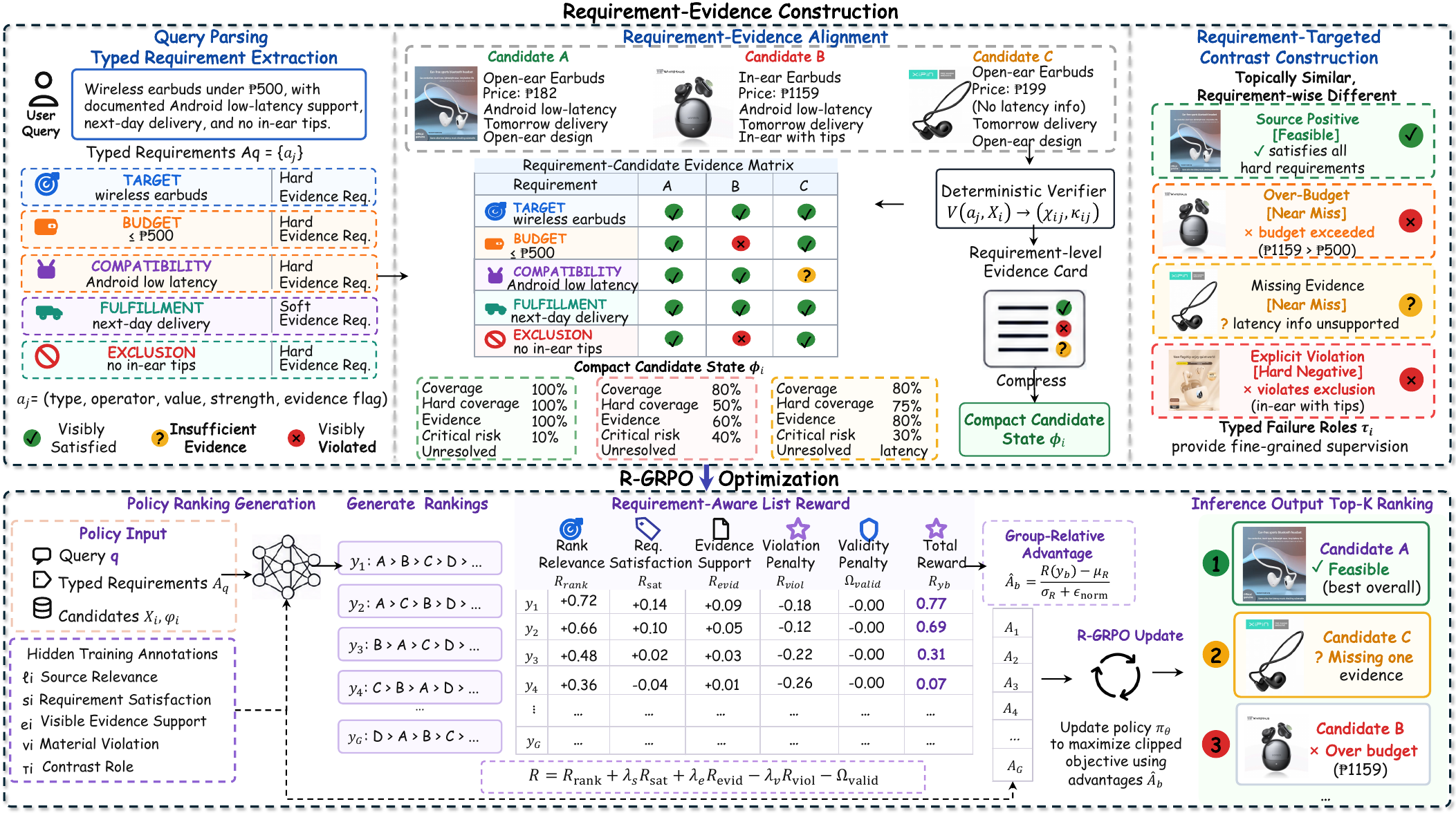}
\caption{
Overview of REAlign. A compositional query is parsed into typed
requirements, visible candidate records are aligned with those requirements
to derive compact candidate states and requirement-targeted contrast roles,
and R-GRPO optimizes partial rankings using decomposed relevance,
requirement-satisfaction, evidence-support, violation, and validity signals.
}
\label{fig:accord-overview}
\end{figure*}

REAlign represents queries and candidates in a shared
requirement--evidence space: typed requirements specify what should
hold, and visible candidate records indicate what is supported,
violated, or unresolved.
This shared representation drives requirement-targeted contrast construction
and requirement-aware partial-list optimization throughout the framework;
Figure~\ref{fig:accord-overview} illustrates the pipeline.

\subsection{Problem Setup}
\label{sec:problem-setup}

Let $q$ denote a compositional query and let
$\mathcal{C}_q=\{d_1,\ldots,d_n\}$ be a bounded candidate pool produced
by an upstream retrieval system.
The reranker then returns an ordered, duplicate-free candidate subset
\begin{equation}
    y=(d_{\pi_1},\ldots,d_{\pi_L}),
    \qquad
    L\leq K,\quad d_{\pi_k}\in\mathcal{C}_q ,
    \label{eq:partial-ranking}
\end{equation}
rather than a permutation of the complete pool, reflecting practical
deployment settings where only the highest-ranked candidates are ultimately exposed.

Source relevance remains the primary signal but cannot order topically
similar candidates; REAlign augments it with requirement satisfaction,
visible evidence support, and material violations.
The fixed-pool setting isolates this ordering problem from first-stage
candidate recall: the method can prevent a near miss from displacing a
more feasible alternative, but it cannot recover an item absent from
$\mathcal{C}_q$.

\subsection{Requirement--Evidence Representation}
\label{sec:req-evidence-representation}

\paragraph{Typed query requirements.}
A compositional query is normalized into a set of typed requirement atoms,
\begin{equation}
    \mathcal{A}_q=\{a_j\}_{j=1}^{m},
    \qquad
    a_j=(t_j,o_j,\nu_j,p_j,r_j),
    \label{eq:requirement-atoms}
\end{equation}
where $t_j$ denotes the requirement type, $o_j$ its operator,
$\nu_j$ its normalized value, $p_j$ its strength, and $r_j$ indicates
whether explicit evidence is required for verification.

The type space covers target concepts, attributes, budgets, exclusions,
compatibility, usage scenarios, and fulfillment conditions.
The strength variable distinguishes hard requirements, whose explicit
violation should strongly affect the ordering, from soft preferences
that can be traded against other useful properties.
The evidence flag is orthogonal to requirement strength: it specifies
whether the model may treat a condition as supported only when a
corresponding catalog field or text span is visible.
Although the offline construction process may organize requirements as
a Query Requirement Graph (QRG), the policy consumes only
the serialized typed records in $\mathcal{A}_q$.
\paragraph{Visible candidate evidence.}
For candidate $d_i$, let $X_i$ denote its visible catalog record
(title, category, price, attributes, fulfillment fields, and short
evidence text).
A deterministic verifier maps each requirement--record pair to a
categorical outcome and a confidence score:
\begin{equation}
    (\chi_{ij},\kappa_{ij})
    =
    \mathcal{V}(a_j,X_i),
    \qquad
    \kappa_{ij}\in[0,1].
    \label{eq:verification-state}
\end{equation}
The outcome
$\chi_{ij}\in
\{\textsc{Satisfied},\textsc{Violated},\textsc{Unsupported}\}$
marks the requirement as supported, contradicted, or unresolved by the
visible record, so missing catalog information is never treated as
negative evidence: a displayed price above budget is a violation,
whereas an absent compatibility field merely leaves the requirement
unsupported.

During construction, an internal Item Evidence Card (IEC) retains the
requirement-linked spans, their sources, and verifier confidence.
We compress the verifier outputs into an evidence-grounded
candidate state
\begin{equation}
    \boldsymbol{\phi}_i
    =
    \left(
        c_i^{\mathrm{all}},
        c_i^{\mathrm{req}},
        c_i^{\mathrm{hard}},
        \bar{\kappa}_i,
        \rho_i,
        b_i,
        \mu_i
    \right).
    \label{eq:candidate-state}
\end{equation}
The three coverage terms summarize the proportions of all requirements,
evidence-required requirements, and hard requirements whose status can
be resolved from the visible record.
$\bar{\kappa}_i$ is the mean confidence over resolved requirements,
while $\rho_i$ is a composite risk score that grows with uncovered
evidence-required requirements, uncovered hard requirements, and low
verifier confidence (its exact form is given in the supplement).
The binary variable $b_i$ indicates whether the candidate has no
uncovered hard evidence-required requirement and low overall risk, and
$\mu_i$ provides a compact summary of the remaining unsupported
conditions.

All components of $\boldsymbol{\phi}_i$ are computed deterministically
from the verifier outcomes $\{(\chi_{ij},\kappa_{ij})\}_{j=1}^{m}$;
neither relevance labels nor contrast roles enter the aggregation.

\subsection{Requirement-Targeted Candidate Contrasts}
\label{sec:requirement-contrasts}

Because candidates with the same relevance label may violate different
constraints or differ in evidence support, similarity-based hard
negatives do not by themselves reveal which requirement makes a
candidate lose.
For each candidate $d_i$, REAlign retains the training signals
\begin{equation}
    \boldsymbol{z}_i
    =
    (\ell_i,s_i,e_i,v_i,\tau_i),
    \label{eq:training-signals}
\end{equation}
where $\ell_i$ is the source relevance label,
$s_i$ and $e_i$ denote requirement satisfaction and visible evidence
support, $v_i$ indicates a material violation, and $\tau_i$ specifies the
candidate role.
The role set distinguishes source positives, semantic hard negatives,
wrong-category items, over-budget items, missing-evidence cases, and
explicit constraint violations.

Training groups are constructed to contain topically similar candidates that differ along identifiable requirements. For example, one candidate may match the requested category and attributes but exceed the budget, while another may satisfy the visible constraints yet lack evidence for compatibility. We refer to these examples as requirement contrasts. The construction is label-preserving: $\ell_i$ remains the sole authority for positive relevance, and neither the candidate role nor the evidence annotation can alter it.

\subsection{Requirement-Aware Partial-List Utility}
\label{sec:partial-list-utility}

The policy autoregressively generates candidate identifiers, and each
output is parsed into a canonical partial list; malformed outputs,
duplicates, and out-of-pool identifiers count as structural errors.
Each ranking is evaluated by a decomposed partial-list utility that
preserves source relevance while explicitly accounting for
requirement-related candidate properties.
For a decoded ranking
$y=(d_{\pi_1},\ldots,d_{\pi_L})$, we define the top-$L$ aggregation of a
candidate-level channel $c$ as
\begin{equation}
    R_c(y)
    =
    \frac{1}{L}
    \sum_{k=1}^{L} c_{\pi_k},
    \label{eq:channel-aggregation}
\end{equation}
a uniform average over the returned partial list.
Setting $c=s$, $e$, and $v$ yields the requirement-satisfaction,
evidential support, and violation terms
$R_{\mathrm{sat}}$, $R_{\mathrm{evid}}$, and
$R_{\mathrm{viol}}$; credit and penalty thus accrue only for candidates
exposed in the partial list.
The base ranking component $R_{\mathrm{rank}}(y)$ uses the frozen
source relevance labels for NDCG, positive coverage, and pairwise
ordering with a label-independent list-coverage term.
Requirement satisfaction, evidence support, material violations, and
candidate roles do not directly enter this ranking component.
The complete partial-list utility preserves source relevance as the
primary ranking objective and explicitly augments it with requirement-aware signals:
\begin{equation}
    R(y)
    =
    R_{\mathrm{rank}}(y)
    +
    \boldsymbol{\lambda}^{\top}
    \mathbf{r}_{\mathrm{req}}(y)
    -
    \Omega_{\mathrm{valid}}(y),
    \label{eq:accord-reward}
\end{equation}
where
$\mathbf{r}_{\mathrm{req}}(y)
=
\bigl(
R_{\mathrm{sat}}(y),
R_{\mathrm{evid}}(y),
-R_{\mathrm{viol}}(y)
\bigr)$
collects requirement satisfaction, evidence support, and material
violations, weighted by $\boldsymbol{\lambda}$. The coefficients remain
fixed and govern the balance among satisfaction, visible evidence, and
explicit violation avoidance throughout policy optimization.
Malformed outputs and missing rankings receive fixed fail-closed
sentinel rewards in place of Equation~\eqref{eq:accord-reward}.
For decodable rankings, let $N_{\mathrm{out}}(y)$ and
$N_{\mathrm{dup}}(y)$ denote the fractions of out-of-pool and duplicate
identifiers and $I_{\mathrm{first}}(y)$ indicate an out-of-pool first
position; the structural penalty is
\begin{equation}
    \Omega_{\mathrm{valid}}(y)
    =
    \gamma_{\mathrm{out}} N_{\mathrm{out}}
    +\gamma_{\mathrm{first}} I_{\mathrm{first}}
    +\gamma_{\mathrm{dup}} N_{\mathrm{dup}},
    \label{eq:validity-penalty}
\end{equation}
with all coefficients fixed throughout training.

The relevance-only GRPO control uses
$R_{\mathrm{rank}}(y)-\Omega_{\mathrm{valid}}(y)$ and removes all
requirement-aware reward channels beyond $R_{\mathrm{rank}}$. GRPO and
R-GRPO start from the same requirement-aware SFT initialization and share the
same policy input, sampling configuration, parser, group-relative
update, and policy-update budget. Their comparison therefore isolates
the incremental contribution of the explicit requirement-aware
reward channels.
\subsection{Optimization with R-GRPO}
\label{sec:rgrpo}

We optimize the partial-list policy using
Requirement-Aware Group-Relative Policy Optimization
(R-GRPO).
For each policy input
$\mathcal{I}_q=
(q,\mathcal{A}_q,\{(X_i,\boldsymbol{\phi}_i)\}_{i=1}^{n})$,
the behavior policy samples a group of $G$ partial rankings:
\begin{equation}
    y_b
    \sim
    \pi_{\theta_{\mathrm{old}}}
    \bigl(\,\cdot\mid\mathcal{I}_q\bigr),
    \qquad
    b=1,\ldots,G .
    \label{eq:group-sampling}
\end{equation}
All samples in the group share the same query, typed requirements, and
candidate pool.
Each sampled ranking is canonicalized and evaluated using
Equation~\eqref{eq:accord-reward}.
Let $\mu_R$ and $\sigma_R$ denote the mean and standard deviation of
the rewards within the group.
The normalized advantage of $y_b$ is
\begin{equation}
    \hat{A}_b
    =
    \frac{R(y_b)-\mu_R}
         {\sigma_R+\epsilon_{\mathrm{norm}}}.
    \label{eq:group-advantage}
\end{equation}
This normalization removes query-specific reward-scale variation within each sampled group.
For each $b\in\{1,\ldots,G\}$ and generated token $y_{b,t}$, the
current-to-behavior importance ratio is
\begin{equation}
    \rho_{b,t}(\theta)
    =
    \frac{
        \pi_{\theta}
        (y_{b,t}\mid y_{b,<t},\mathcal{I}_q)
    }{
        \pi_{\theta_{\mathrm{old}}}
        (y_{b,t}\mid y_{b,<t},\mathcal{I}_q)
    }.
    \label{eq:token-ratio}
\end{equation}
Let
$\mathcal{M}_b=\{t:m_{b,t}=1\}$
denote the set of optimized token positions for completion $y_b$,
excluding padding and other masked positions.
Using the standard clipped surrogate
\begin{equation}
\scalebox{0.90}{$\displaystyle
    \mathcal{C}_{\epsilon_{\mathrm{clip}}}(\rho,A)
    =
    \min
    \left\{
        \rho A,\,
        \operatorname{clip}(\rho,1-\epsilon_{\mathrm{clip}},1+\epsilon_{\mathrm{clip}})A
    \right\},
$}
    \label{eq:clipped-surrogate}
\end{equation}
the policy objective is
\begin{equation}
    \mathcal{L}_{\mathrm{R}}
    =
    -\frac{1}{G}
    \sum_{b=1}^{G}
    \left[
        \frac{1}{|\mathcal{M}_b|}
        \sum_{t\in\mathcal{M}_b}
        \mathcal{C}_{\epsilon_{\mathrm{clip}}}
        \bigl(
            \rho_{b,t}(\theta),
            \hat{A}_b
        \bigr)
    \right].
    \label{eq:rgrpo-objective}
\end{equation}

R-GRPO retains the standard clipped group-relative update, but derives
its advantage from requirement-aware list utility rather than relevance
alone during policy optimization: rankings with comparable topical relevance
receive different learning signals when they differ in requirement satisfaction,
visible evidence support, or material violations.
\subsection{Training and Inference}
\label{sec:training-inference}

During training, the policy input consists of the original query,
the serialized requirement set, and the visible candidate records
augmented with their compact evidence summaries for grounded listwise
policy optimization:
\begin{equation}
\left(
    q,\mathcal{A}_q,
    \left\{
        \left(X_i,\boldsymbol{\phi}_i\right)
    \right\}_{i=1}^{n}
\right).
\label{eq:policy-input}
\end{equation}
The source relevance, satisfaction, evidence, violation, and role
annotations in $\boldsymbol{z}_i$ are kept outside the prompt and are
used only to evaluate sampled rankings; the policy never reads the
labels that determine its reward.
At inference time, the training-only channels are absent: the model
receives the same requirement and visible-field input and generates a
machine-readable, duplicate-free partial top-$K$ ordering.
\section{Experiments}
We evaluate REAlign through six research questions.
\textbf{RQ1:} How does R-GRPO compare with the supervised fine-tuning
(SFT) initialization and policy-optimization baselines in fixed-pool
reranking?
\textbf{RQ2:} Do its ranking gains translate into improved requirement
satisfaction, evidence support, and fewer violations?
\textbf{RQ3:} What are the contributions of the requirement and evidence
representations?
\textbf{RQ4:} How does each reward component affect ranking behavior?
\textbf{RQ5:} How does performance vary with query complexity?
\textbf{RQ6:} How consistent are the gains across requirement types?
\subsection{Experimental Setup}
\label{sec:experimental-setup}

\paragraph{Benchmarks.}
We evaluate REAlign on two fixed-pool e-commerce reranking benchmarks.
\textsc{Shop-Need} is derived from ShoppingBench
\citep{wang2026shoppingbench} and contains 4,880 query groups, each with
12 candidates.
\textsc{KS-Need} is derived from KuaiSearch
\citep{li2026kuaisearch} and contains 5,000 groups grounded in
real search interactions.
Both datasets introduce compositional requirements such as budget,
attributes, exclusions, compatibility, and fulfillment, while preserving
relevance labels inherited from their source collections.

\begin{table}[!t]
\centering
\begin{minipage}{\linewidth}
\centering
\setlength{\tabcolsep}{9.3pt}
\begin{tabular}{lrrrr}
\toprule
Dataset & Groups & Train & Test & Pairs \\
\midrule
Shop-Need & 4,880 & 3,904 & 976 & 58,560 \\
KS-Need & 5,000 & 4,000 & 1,000 & 59,478 \\
\bottomrule
\end{tabular}
\end{minipage}
\caption{
Statistics of the two fixed-pool reranking benchmarks.
Groups and Pairs denote query groups and query--candidate pairs, respectively, used in our experiments.
}
\label{tab:data}

\end{table}

\begin{table*}[!t]
\centering
\footnotesize
\setlength{\tabcolsep}{2pt}
\renewcommand{\arraystretch}{1.08}
\begin{tabular*}{\linewidth}{@{\extracolsep{\fill}}llrrrrrrrrrr@{}}
\toprule
& & \multicolumn{6}{c}{Ranking Quality} & \multicolumn{4}{c}{Diagnostics} \\
\cmidrule(lr){3-8}\cmidrule(lr){9-12}
Dataset & Method & N@5 & N@10 & H@1 & H@5 & H@10 & Pars. & R@5 & V@5 $\downarrow$ & E@5 & Rew@5 \\
\midrule
\multirow{8}{*}{Shop-Need} & SFT & 0.5891 & 0.6031 & 0.5830 & 0.6650 & 0.6957 & \textbf{1.0000} & 0.0652 & 0.7084 & 0.0670 & 0.2102 \\
\addlinespace[1pt]
 & PPO~\citep{schulman2017proximal} & 0.5904 & 0.6070 & 0.5799 & 0.6650 & 0.7039 & 0.9980 & 0.0643 & 0.7057 & 0.0662 & 0.2098 \\
 & DPO~\citep{rafailov2023direct} & 0.5917 & 0.6053 & 0.5809 & 0.6701 & 0.6988 & 0.9980 & 0.0656 & 0.7082 & 0.0676 & 0.2111 \\
 & GRPO~\citep{shao2024deepseekmath} & 0.8117 & 0.8153 & 0.8484 & 0.8617 & 0.8658 & \textbf{1.0000} & 0.0793 & 0.6752 & 0.0814 & 0.3110 \\
 & DAPO~\citep{yu2025dapo} & 0.8748 & 0.8795 & 0.9139 & 0.9273 & 0.9314 & \textbf{1.0000} & 0.0822 & 0.6633 & 0.0842 & 0.3381 \\
 & GSPO~\citep{zheng2025gspo} & 0.9131 & 0.9195 & 0.9436 & 0.9621 & 0.9662 & \textbf{1.0000} & 0.0846 & 0.6553 & 0.0865 & 0.3528 \\
 & GDPO~\citep{liu2026gdpo} & 0.9071 & 0.9129 & 0.9436 & 0.9529 & 0.9549 & \textbf{1.0000} & 0.0848 & 0.6566 & 0.0867 & 0.3515 \\
 & Full R-GRPO & \textbf{0.9226} & \textbf{0.9293} & \textbf{0.9488} & \textbf{0.9693} & \textbf{0.9734} & 0.9969 & \textbf{0.0852} & \textbf{0.6535} & \textbf{0.0873} & \textbf{0.3561} \\
\midrule
\multirow{8}{*}{KS-Need} & SFT & 0.2456 & 0.4050 & 0.1570 & 0.5540 & 0.8970 & 0.9700 & 0.1320 & 0.0958 & 0.7008 & 0.3782 \\
\addlinespace[1pt]
 & PPO~\citep{schulman2017proximal} & 0.2404 & 0.4034 & 0.1530 & 0.5520 & 0.8950 & 0.9630 & 0.1298 & 0.0956 & 0.7010 & 0.3763 \\
 & DPO~\citep{rafailov2023direct} & 0.2064 & 0.3309 & 0.1400 & 0.4240 & 0.7940 & 0.7100 & 0.1070 & 0.1194 & 0.6690 & 0.3541 \\
 & GRPO~\citep{shao2024deepseekmath} & 0.3520 & 0.5049 & 0.2570 & 0.7150 & 0.9510 & \textbf{1.0000} & 0.1784 & 0.0946 & 0.6888 & 0.4203 \\
 & DAPO~\citep{yu2025dapo} & 0.2981 & 0.4043 & 0.2020 & 0.6560 & 0.8760 & \textbf{1.0000} & 0.1604 & 0.0958 & 0.6938 & 0.3990 \\
 & GSPO~\citep{zheng2025gspo} & 0.3269 & 0.4908 & 0.2370 & 0.6770 & \textbf{0.9580} & \textbf{1.0000} & 0.1636 & 0.0872 & 0.7070 & 0.4145 \\
 & GDPO~\citep{liu2026gdpo} & 0.3192 & 0.4704 & 0.2230 & 0.6820 & 0.9390 & \textbf{1.0000} & 0.1632 & 0.0826 & \textbf{0.7152} & 0.4146 \\
 & Full R-GRPO & \textbf{0.3891} & \textbf{0.5173} & \textbf{0.2970} & \textbf{0.7750} & 0.9530 & \textbf{1.0000} & \textbf{0.1894} & \textbf{0.0792} & 0.7138 & \textbf{0.4441} \\
\bottomrule
\end{tabular*}
\caption{Held-out fixed-pool results for RQ1--RQ2. N, H, R, V, E, Rew, and Pars. denote NDCG, HR, ReqSat, Violation, Evidence, the composite reward, and recoverable in-pool ranking output. Each SFT row evaluates the supervised initialization; every policy-optimization row starts from that adapter. Bold marks best value per dataset; lower is better only for V@5.}
\label{tab:main-ranking}
\end{table*}
\paragraph{Baselines.}
We compare against GRPO, DAPO, GSPO, GDPO, PPO, and DPO, spanning
group-relative, sequence-level, actor--critic, and preference-based
optimization. The full requirement-aware REAlign model is denoted
Full R-GRPO in tables and R-GRPO elsewhere.

\paragraph{Implementation Details.}
All methods use the same Qwen3.5-4B backbone~\citep{qwenteam2026qwen35} and dataset-specific SFT
initialization across all experiments. Policy-optimization methods share
the reranking interface, frozen evaluator, and matched update budget;
held-out labels are excluded from training and checkpoint selection.
Supplementary Appendix D.4 provides optimizer-specific implementations and
hyperparameters.

\paragraph{Metrics.}
We report NDCG@K and HR@K on held-out groups using source relevance
labels unavailable during training and checkpoint selection for those groups.
We also report ReqSat@5, Evidence@5, and Violation@5 to more directly characterize how
the generated rankings behave with respect to requirement annotations.
Reward@5 is reported as an optimization sanity check, measuring how well each
method realizes the predefined composite objective on held-out groups. Pars.
denotes the fraction of recoverable in-pool rankings produced by each method.
\subsection{RQ1: Overall Ranking Performance}
\label{sec:rq1}

Table~\ref{tab:main-ranking} shows that REAlign consistently delivers the
strongest overall ranking performance on both benchmarks under matched
backbone, initialization, and optimization budgets.
On \textsc{Shop-Need}, R-GRPO outperforms the SFT initialization and all
policy-optimization baselines on every NDCG and hit-rate metric.
On \textsc{KS-Need}, it achieves the best NDCG and early-rank hit rates;
on HR@10 and Evidence@5, a paired bootstrap over held-out groups
does not resolve a difference from the best baseline.
The baselines also separate cleanly: PPO barely moves from the
supervised initialization and DPO can even degrade it, whereas
group-relative methods improve substantially, with the
requirement-aware objective adding a further margin over the strongest
of them.
The gains concentrate most clearly at shallow cutoffs, and the consistent
advantage over relevance-only GRPO indicates that the improvement arises
from requirement-aware supervision rather than the group-relative update
alone.

\subsection{RQ2: Requirement-Aware Diagnostics}
\label{sec:rq2}

Table~\ref{tab:main-ranking} shows that the ranking improvements are
accompanied by the intended requirement-aware behavior. Full R-GRPO
achieves the lowest Violation@5 and the highest ReqSat@5 and Reward@5
on both benchmarks, while maintaining competitive evidence support.
Because these diagnostics are derived from the frozen construction
annotations used to define the reward, they measure objective-aligned
behavior on held-out groups rather than independent human judgments.
Together with the consistent gains in NDCG and HR, these results suggest
that R-GRPO more effectively prioritizes annotated feasible candidates
over topically relevant near misses without compromising source relevance.
\subsection{RQ3: Representation Ablation}
\label{sec:rq3}

Table~\ref{tab:rq3-representation} evaluates the query requirement
graph (QRG) and candidate-side item evidence card (IEC) under otherwise
identical training and evaluation settings, isolating their respective
contributions more clearly. The full model leads on all six metrics across
datasets, showing that both representations contribute to
requirement-aware reranking. Their effects differ by dataset.
On \textsc{Shop-Need}, removing IEC causes a larger degradation than
removing QRG, highlighting candidate-side evidence grounding. On
\textsc{KS-Need}, QRG has a larger effect on ranking quality, while IEC
remains important for evidence-related diagnostics. Removing both
representations produces the weakest results on \textsc{KS-Need} and
remains well below the full model on \textsc{Shop-Need}. These results support a complementary division of labor, with QRG organizing
compositional constraints and IEC directly grounding their evaluation in
visible candidate evidence.
\begin{table}[!ht]
\centering
\small
\setlength{\tabcolsep}{1pt}
\scalebox{0.92776}{%
\begin{tabular}{llrrrrrrr}
\toprule
Dataset & Variant & N@10 & H@10 & R@5 & V@5 $\downarrow$ & E@5 & Rew@5 & Pars. \\
\midrule
\multirow{4}{*}{\shortstack{Shop-\\Need}}
 & Ours & \textbf{0.9293} & \textbf{0.9734} & \textbf{0.0852} & \textbf{0.6535} & \textbf{0.0873} & \textbf{0.3561} & 0.9969 \\
 & w/o QRG & 0.9067 & 0.9611 & 0.0834 & 0.6583 & 0.0852 & 0.3420 & 1.0000 \\
 & w/o IEC & 0.8213 & 0.9201 & 0.0750 & 0.6605 & 0.0762 & 0.3009 & 1.0000 \\
 & w/o both & 0.8814 & 0.9457 & 0.0766 & 0.6618 & 0.0779 & 0.3340 & 1.0000 \\
\midrule
\multirow{4}{*}{\shortstack{KS-\\Need}}
 & Ours & \textbf{0.5173} & \textbf{0.9530} & \textbf{0.1894} & \textbf{0.0792} & \textbf{0.7138} & \textbf{0.4441} & \textbf{1.0000} \\
 & w/o QRG & 0.4828 & 0.9480 & 0.1718 & 0.0810 & 0.7034 & 0.4169 & 0.9940 \\
 & w/o IEC & 0.5086 & 0.9518 & 0.1726 & 0.0802 & 0.7017 & 0.4152 & 1.0000 \\
 & w/o both & 0.4736 & 0.9352 & 0.1695 & 0.0824 & 0.6950 & 0.4039 & 1.0000 \\
\bottomrule
\end{tabular}%
}
\caption{
Ablation of query requirement graph (QRG) and
candidate-side item evidence card (IEC).
Lower is better for V@5.
Bold denotes the best within each dataset for each of the six metrics.
Pars. reports the fraction of recoverable in-pool rankings and is excluded
from the comparison.
}
\label{tab:rq3-representation}
\end{table}

\begin{table}[!htbp]
\centering
\scriptsize
\setlength{\tabcolsep}{1pt}
\scalebox{1.13272}{%
\begin{tabular}{llrrrrrrr}
\toprule
Dataset & Reward & N@10 & H@10 & R@5 & V@5 $\downarrow$ & E@5 & Rew@5 & Pars. \\
\midrule
\multirow{6}{*}{\shortstack{Shop-\\Need}}  & GRPO & 0.8153 & 0.8658 & 0.0793 & 0.6752 & 0.0814 & 0.3110 & \textbf{1.0000} \\
 & w/o $R_{\mathrm{sat}}$ & 0.8262 & 0.8791 & 0.0807 & 0.6717 & 0.0828 & 0.3159 & \textbf{1.0000} \\
 & w/o $R_{\mathrm{evid}}$ & 0.9114 & 0.9590 & 0.0832 & 0.6539 & 0.0852 & 0.3471 & 0.9990 \\
 & w/o $R_{\mathrm{viol}}$ & 0.8185 & 0.8730 & 0.0811 & 0.6732 & 0.0832 & 0.3106 & \textbf{1.0000} \\
 & w/o $\Omega_{\mathrm{valid}}$ & 0.9018 & 0.9518 & 0.0836 & 0.6594 & 0.0857 & 0.3431 & \textbf{1.0000} \\
 & R-GRPO & \textbf{0.9293} & \textbf{0.9734} & \textbf{0.0852} & \textbf{0.6535} & \textbf{0.0873} & \textbf{0.3561} & 0.9969 \\
\midrule
\multirow{6}{*}{\shortstack{KS-\\Need}} & GRPO & 0.5049 & 0.9510 & 0.1784 & 0.0946 & 0.6888 & 0.4203 & \textbf{1.0000} \\
 & w/o $R_{\mathrm{sat}}$ & 0.4964 & 0.9410 & 0.1748 & 0.0838 & \textbf{0.7198} & 0.4273 & \textbf{1.0000} \\
 & w/o $R_{\mathrm{evid}}$ & 0.4942 & 0.9250 & \textbf{0.1910} & 0.0842 & 0.7026 & 0.4430 & \textbf{1.0000} \\
 & w/o $R_{\mathrm{viol}}$ & 0.4929 & 0.9490 & 0.1720 & 0.0918 & 0.6984 & 0.4142 & \textbf{1.0000} \\
 & w/o $\Omega_{\mathrm{valid}}$ & 0.4947 & 0.9420 & 0.1684 & 0.0818 & 0.7178 & 0.4245 & \textbf{1.0000} \\
 & R-GRPO & \textbf{0.5173} & \textbf{0.9530} & 0.1894 & \textbf{0.0792} & 0.7138 & \textbf{0.4441} & \textbf{1.0000} \\
\bottomrule
\end{tabular}%
}
\caption{
Reward ablation with the relevance-only GRPO baseline, single-channel
deletions, and the full requirement-aware objective.
Lower is better for V@5.
}
\label{tab:reward-components}
\end{table}

\subsection{RQ4: Reward Ablation}
\label{sec:rq4}

Table~\ref{tab:reward-components} compares the relevance-only GRPO
baseline with single-channel deletions from the requirement-aware
objective. All variants start from the same SFT initialization and are
trained under otherwise fixed settings. The full objective achieves the
highest Reward@5 on both benchmarks, supporting the contribution of its
reward channels in this setting for compositional reranking. On
\textsc{Shop-Need}, removing the satisfaction or violation channel
eliminates most of the gain over GRPO, whereas removing the evidence
channel has a smaller effect. On \textsc{KS-Need}, every single-channel
deletion reduces NDCG@10 below GRPO, suggesting that an incomplete
requirement-aware objective can underperform its relevance-only
counterpart. The channels also induce trade-offs: removing
$R_{\mathrm{sat}}$ yields the highest Evidence@5 on \textsc{KS-Need} at
the expense of ranking quality, while $R_{\mathrm{viol}}$ contributes
most strongly to reducing Violation@5.

\subsection{RQ5: Performance across Query Complexity}
\label{sec:rq5}

\begin{figure}[!t]
\centering
\includegraphics[width=\columnwidth]{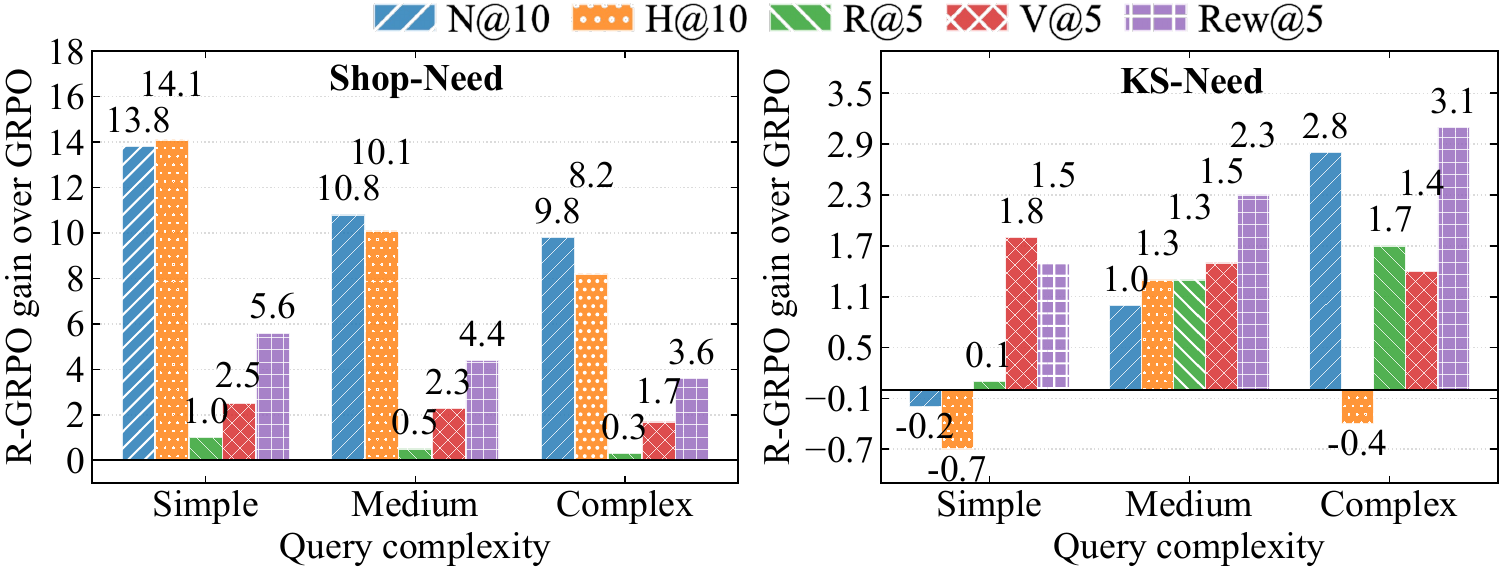}
\caption{
R-GRPO gains over GRPO across query-complexity strata on the held-out
sets, measured in percentage points. Positive V@5 values indicate fewer violations.
}
\label{fig:complexity-gains}
\end{figure}

Figure~\ref{fig:complexity-gains} compares R-GRPO with GRPO across
Simple, Medium, and Complex query groups.
On \textsc{Shop-Need}, R-GRPO improves all reported metrics at every
complexity level; on \textsc{KS-Need}, the N@10 and Rew@5 gains grow
with complexity and V@5 improves throughout, despite minor H@10
fluctuations.
The \textsc{KS-Need} gains are near zero for Simple queries, where
relevance already determines the ordering, and largest for Complex
ones, where the feasible subset narrows and the ways to be a near miss
multiply, which is precisely where requirement-aware optimization
matters.

\begin{figure}[!htbp]
\centering
\includegraphics[width=\columnwidth]{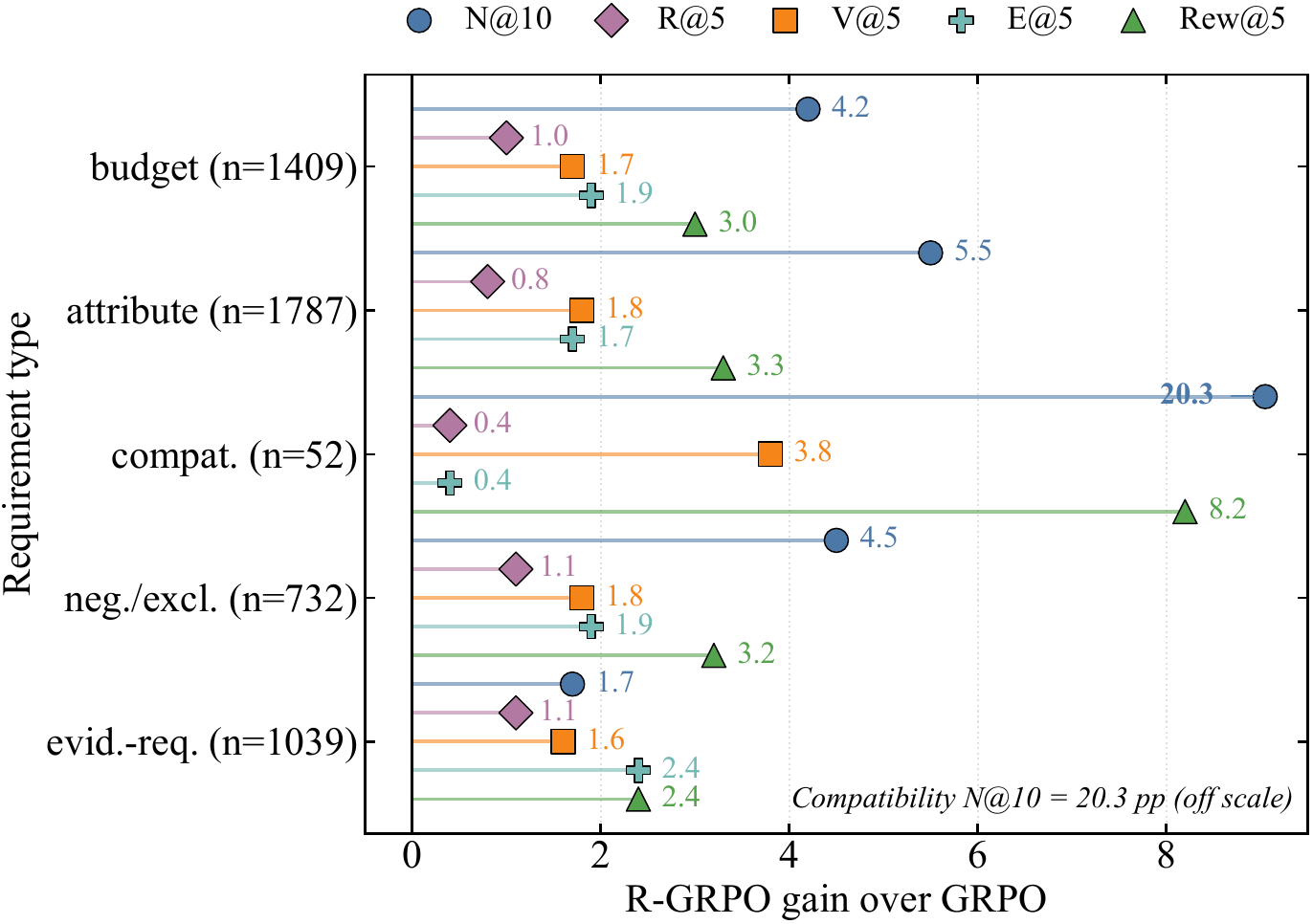}
\caption{
Performance gains of R-GRPO over GRPO across requirement types on
held-out sets, measured in percentage points. Requirement slices are
non-exclusive, with support shown in parentheses. Positive V@5 indicates
fewer violations. The compatibility result is shown off scale and should
be interpreted cautiously given its limited support ($n=52$).
}
\label{fig:requirement-type-gains}
\end{figure}
\subsection{RQ6: Performance across Requirement Types}
\label{sec:rq6}
Figure~\ref{fig:requirement-type-gains} groups the pooled held-out queries
by requirement type, with overlapping subsets.
R-GRPO improves every reported metric across all types, with consistent
gains throughout. The improvements remain consistent despite substantial
variation in subset size and requirement frequency across categories.
Among the well-supported slices, budget, attribute, and
negation/exclusion yield N@10 gains of 4.2, 5.5, and 4.5 points,
respectively. Compatibility shows a larger 20.3-point gain, but its
support is limited to 52 queries and should be interpreted cautiously.
Evidence-required queries show smaller ranking gains but the largest
improvements in visible support, consistent with RQ4.

\section{Conclusion}
We presented REAlign, coupling typed requirements, evidence-grounded candidate states, requirement-targeted contrasts, and requirement-aware group-relative optimization for compositional e-commerce reranking. Across two fixed-pool benchmarks, REAlign consistently improves held-out ranking quality while moving construction-based requirement diagnostics in the intended direction, including fewer verifier-identified violations at top ranks. Because these diagnostics are derived from frozen annotations used to define the requirement-aware reward, they demonstrate objective-aligned behavioral generalization rather than independently validated user utility. Future work should evaluate the framework using human requirement judgments, natural traffic, and end-to-end retrieval settings.
\bibliography{needrank_refs}

\end{document}